\documentclass[a4paper]{jpconf}
\usepackage{graphicx}

\usepackage[english]{babel}
\usepackage[utf8]{inputenc} 
\usepackage[T1]{fontenc}    
\usepackage{hyperref}       
\usepackage{url}            
\usepackage{booktabs}       
\usepackage{amsfonts}       
\usepackage{nicefrac}       
\usepackage{microtype}      
\usepackage{ dsfont }
\usepackage{chemformula}
\usepackage{siunitx}
\usepackage{subfig}
\usepackage{array}
\usepackage{tabularx}
\usepackage{csquotes}

\newcommand{\nanofiber}{NF}

\begin{document}
\title{Hybrid device for quantum nanophotonics}
\author{S Pierini$^{1,2}$, M D'Amato$^2$, M Joos$^2$, Q Glorieux$^2$, E Giacobino$^2$, E~Lhuillier$^3$, C Couteau$^1$ and A Bramati$^2$}

\address{$^1$ Laboratory Light, nanomaterials \& nanotechnologies – L2n, University of Technology of Troyes \& CNRS ERL 7004, 12 rue Marie Curie, 10000 Troyes, France}
\address{$^2$ Laboratoire Kastler Brossel, Sorbonne Université, Ecole Normale Supérieure and CNRS, 4 place Jussieu, 75252 Paris Cedex 05, France}

\address{$^3$ Sorbonne Université, CNRS - UMR 7588, Institut des NanoSciences de Paris, INSP, F-75005 Paris, France}

\ead{alberto.bramati@lkb.upmc.fr}

\bibliographystyle{bibst/iopart-num} 

\begin{abstract}
Photons have been identified early on as a very good candidate for quantum technologies applications, as carriers of quantum information, either by polarization encoding, time encoding or spatial encoding. Quantum cryptography, quantum communications, quantum networks and quantum computing~\cite{acin2018quantum} are some of the applications targeted by the so called quantum photonics. Nevertheless, it was also clear at an early stage that bulk optics for handling quantum states of light would not be the best option for these technologies~\cite{vamivakas2017special}. More recently, single photons, entangled photons and quantum light in general have been coupled to integrated approaches coming from classical optics in order to meet the requirements of scalability, reliability and efficiency for quantum technologies. In this article, we describe our recent advances using elongated optical nano-fibers. We also present our latest results on nanocrystals made of perovskites and discuss some of their quantum properties. Finally, we will discuss the general steps necessary in order to couple these nanoemitters efficiently with our photonic platform, based on taperd optical nanofibers.
\end{abstract}

\section{Introduction}
For future quantum technologies and in particular for quantum architecture systems, one needs to consider scalability and reproducibility in order to be able to handle and add easily as many quantum bits/qubits of information as possible on a given platform. 
Photons are of particular interest as they are good carriers of quantum information and one aim is to explore a fully integrated photonic quantum circuit. This circuit would be a hybrid system made of stationary solid-state qubits (quantum emitters) coupled together via single photons travelling within a common optical bus.
This optical bus should be photonics-ready i.e. compatible with optical fibers for quantum communications within a network of quantum nodes. 
Actually there are very good single photon sources, like single quantum dots in micropillars, that produce very good single photons but require to work at cryogenic temperature and to collect the photons via a microscope: this two characteristics pose a limit in the scalabilty and usability of the system. 
Single photon generation at room temperature and their compact integration with single mode fiber is the goal of our study.
The first part of this article will describe our latest results on quantum emitters based on perovskite nanocrystals that can be synthetised chemically giving rise to single-photons emitters. In the second part we describe the advances in the realization of a photonic platform, based on the technique of the elongated nanofibers~\cite{black1988tapered} in which, due to the sub-wavelength size of the fiber, its optical mode is mostly outside the fiber itself and the strong evanescent field enhances the interaction with an emitter coupled to the fiber.  
Finally, we will present our preliminary results in coupling quantum emitters with nanofibers making use of this strong near field interaction.

\section{Quantum Emitters}
\subsection{Perovskites nanocrystals}
\begin{figure}
    \centering
    \includegraphics[width=0.7\linewidth]{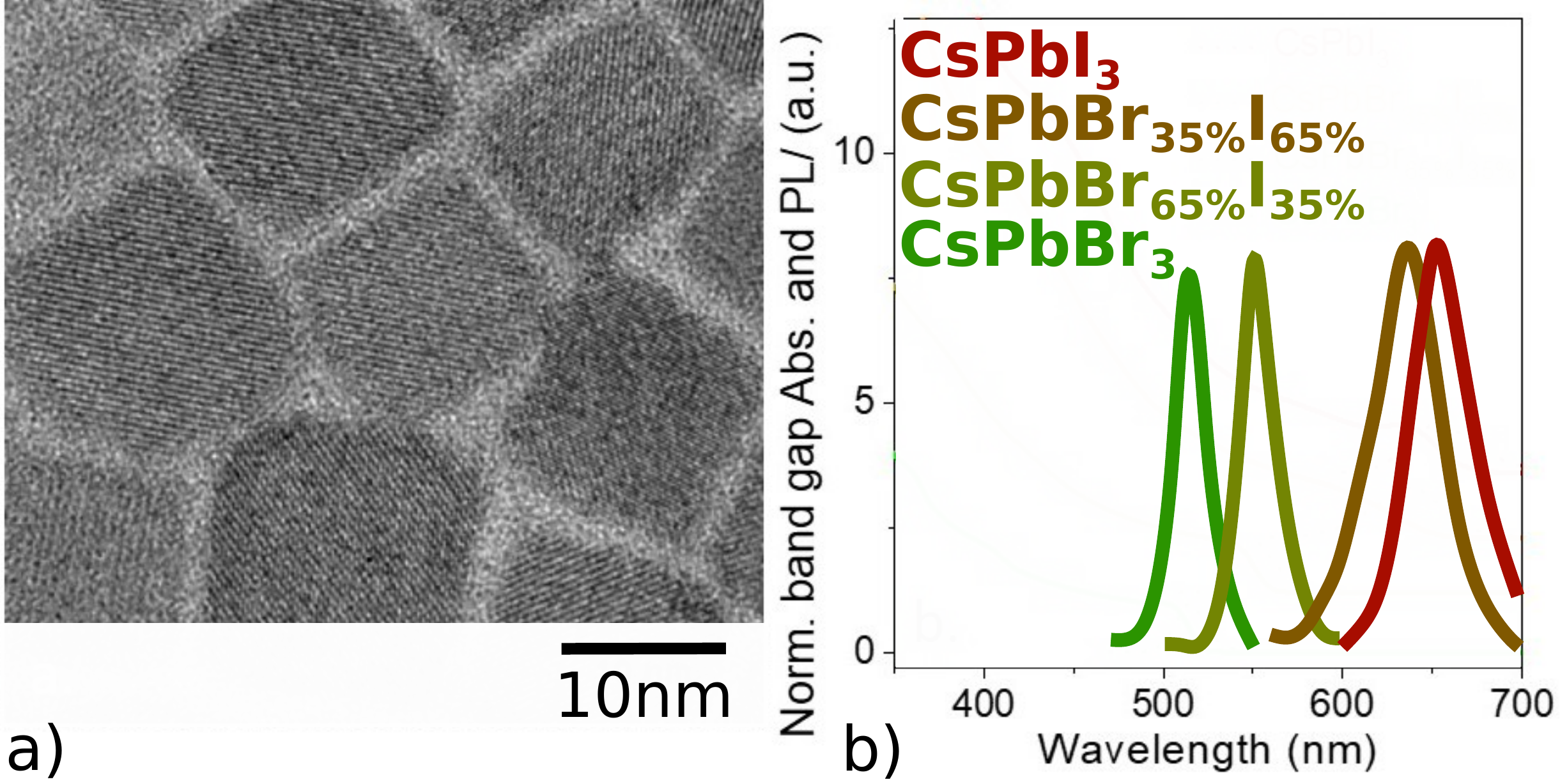}
    \caption{a) transmission electron microscopy image of \ch{CsPb(Br ;I)3} nanocrystal cubes, where the notation denotes a variable ratio between \ch{Br} and \ch{I}. b) Photoluminescence of \ch{CsPb(Br ;I)3} perovskites nanocrystal with various \ch{Br} content.}
    \label{fig:perov1}
\end{figure}

\paragraph{General description}
Nanocrystals are semiconductor nanoparticles with broadly tunable optical features from UV to THz \cite{goubet2018terahertz}. Thanks to quantum confinement, atomic-like spectra can be obtained from such objects \cite{murray1993synthesis}. They thus appear to be an interesting candidate for quantum optics as single photon emission has been reported from them in the past \cite{brokmann2004highly}. Nevertheless,  traditional II-VI semiconductor nanocrystals suffer from several issues. In particular, bright emission is only obtained through the growth of core shell heterostructures. Meanwhile, perovskite nanocrystals made of lead halide materials became very popular in the field of solar cells. Their defect-tolerant electronic structure leads to improved open-circuit voltage. It has been proposed to take advantage of this defect tolerance of lead halide perovskite to design bright core only nanocrystals \cite{Protesescu2015}. 
The basic concept of the synthesis relies on the reaction of lead halide, in presence of long chain ligands made of oleic acid and oleylamine in a non coordinating solvent. At relatively low temperature (180 °C), the injection of a cesium oleate precursor leads to an immediate formation of \ch{CsPbX3} nanocrystals, with \ch{X=Cl}, \ch{Br} or \ch{I}  with cubic shape: this is clearly visible in figure \ref{fig:perov1}a where a transmission electron microscopy image of \ch{CsPb(Br ;I)3} is shown (this formula concisely indicates a structure with different proportion of \ch{Br} and \ch{I}, keeping fixed the proportion between the sum of the moles of \ch{Br} and \ch{I} and the other components of the crystal; this is possible because the two non-metals have the same oxidation number).
These particles have size around \SI{10}{nm}. As for II-VI semiconductor nanocrystals, the band-edge energy can be tuned thanks to quantum confinement. Smaller nanocrystals can simply be obtained by reducing the growth temperature. However this is not the method the most commonly used. Conventional alloying is usually used. We can see in figure~\ref{fig:perov1} the  photoluminescence spectra of different alloying perovskite-nanocrystals in solution leading to different emission wavelengths.
The whole visible range can be spanned by tuning the X from \ch{Cl} to \ch{I}. The most striking property of these perovskite nanocrystals comes from their high photoluminescence efficiency which ranges from $50\%$ to $90\%$ depending on the chosen halide. 

\paragraph{Optical characterization} 
\begin{figure}
    \centering
    \includegraphics[width=0.5\linewidth]{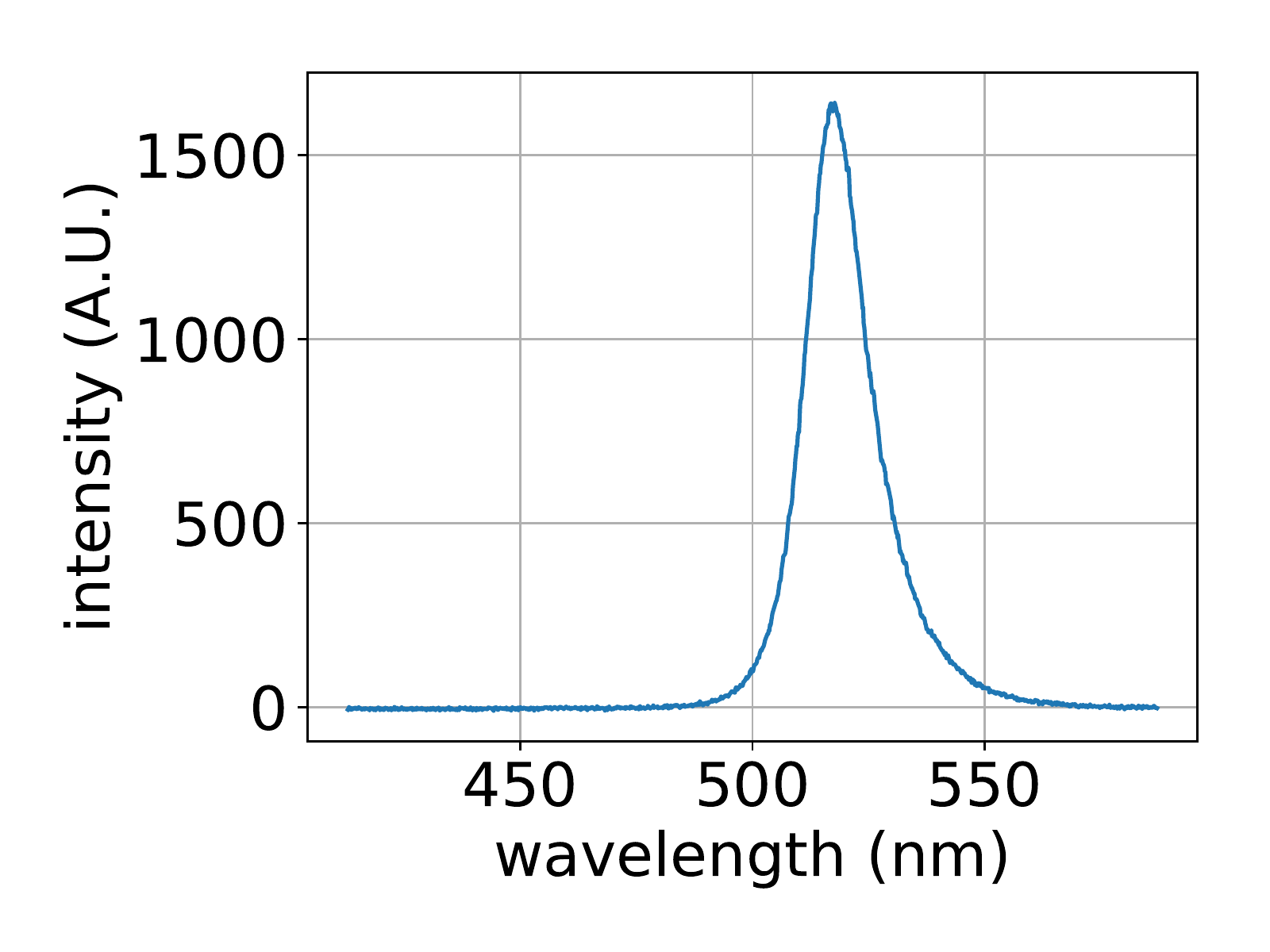}
    \caption{Spectrum of a perovskite nanocrystal with a center wavelength of \SI{518}{nm} and a full width at half maximum \SI{16}{nm}.}
    \label{fig:perov_spectra}
\end{figure}
To study the emission spectra of these nanocrystals, we deposit them on a glass plate and use a confocal microscope to excite a single emitter with a \SI{405}{nm} pulsed excitation laser. The central wavalenghts of the spectra spread from \SI{480}{nm} to \SI{520}{nm} with a peak with a full width at half maximum (FWHM) of about \SI{15}{nm}. A typical spectrum is shown in figure \ref{fig:perov_spectra}.
\begin{figure}
    \centering
    \includegraphics[width=0.5\linewidth]{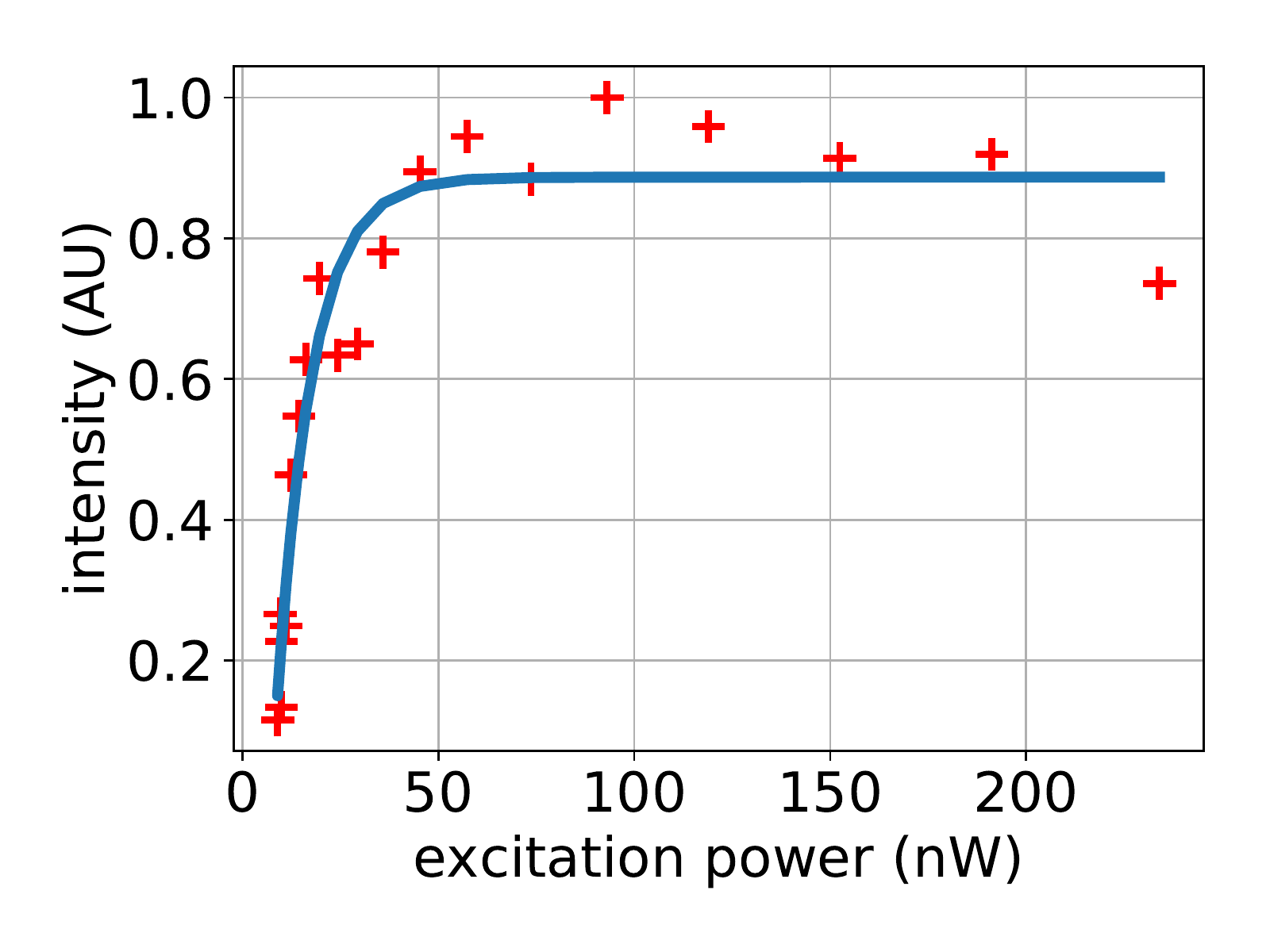}
    \caption{Saturation curve of a perovskite nanocrystal showing single photon emission. The nanocrystal is excited with a Picoquant picosecond pulsed laser with a repetition rate of \SI{200}{\nano\second} and focalised with a \SI{100}{X} Nikon oil-immersion objective with a numerical aperture of \SI{1.4}{}.}
    \label{fig:perov_saturation}
\end{figure}
The emission of a single nanocrystal shows a clear saturation of the emitted light while increasing the excitation power, with a saturation power of \SI{80}{nW} (the curve is shown in Figure~\ref{fig:perov_saturation}). Each experimental point is obtained by taking ten measurements and keeping only the three brightest ones. This procedure allows to select the measurements where the emitters was on and to filter out the effect of the blinking.
The measurements of the Stokes parameters performed on our samples with the method illustrated in
\cite{berry1977measurement} have shown that the emission is not polarized.   

\paragraph{Quantum properties} Antibunching measurements were performed using a Hanbury Brown and Twiss setup showing that some of the emitters had a clear single photon emission, with a $g^{(2)}(0)<0.1$. 
With our setup we can measure $g^{(2)}(\tau)$ 
for long values of $\tau$ (up to hundreds of \si{\milli\second}). This feature allows us to normalize the $g^{(2)}$ function at large delays, taking into account the bunching effect due to the blinking~\cite{manceau2018cdse}. This effect has to be considered in order to identify the quality of the emitter and it is taken in consideration for the first time for perovskite nanocrystals.
\begin{figure}
    \centering
    \includegraphics[width=0.8\linewidth]{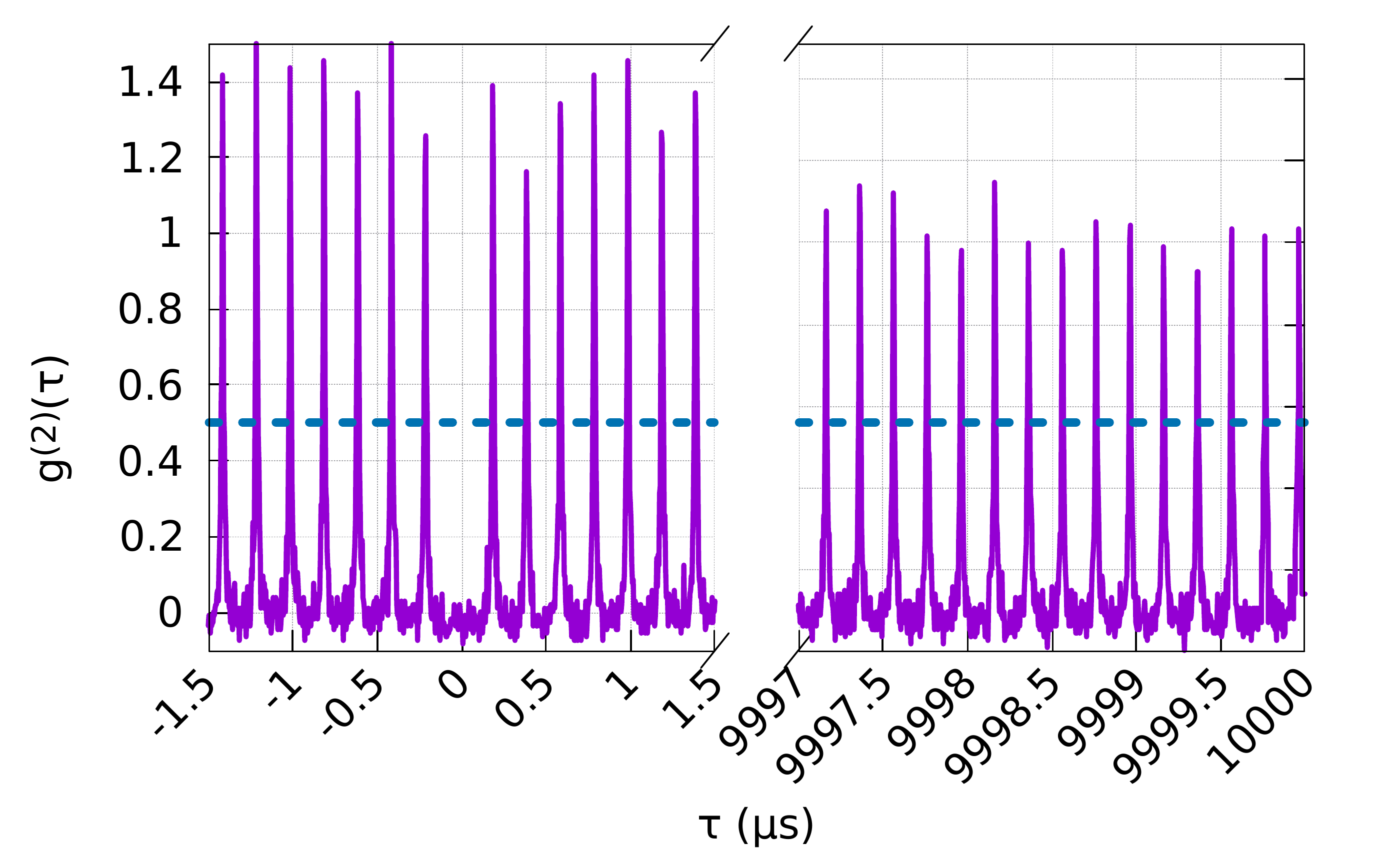}
    \caption{Measurement of  the autocorrelation function $g^{(2)}(\tau)$ of a single perovskite nanocrystal, showing clear single photon emission. The data are corrected for the noise (background removed) and normalized at large delays ($\tau \approx \SI{10}{\milli\second}$). The  $g^{(2)}(\tau)$ is shown around $\tau=0$ and for $\tau \approx \SI{10}{\milli\second}$. The blue line is at $g^{(2)}=0.5$, as $g^{(2)}(0)<0.5$ we have single photon emission.}
    \label{fig:g2_perov}
\end{figure}
The result of the photon antibunching is shown in figure~\ref{fig:g2_perov}. The experimental curves are normalized using the peaks at a long delay
$\tau\approx\SI{100}{\micro\second}$, 
after having subtracted the background noise, estimated using the values between two adjacent peaks. To perform the measurement, we use a PicoHarp~800 with a router that registers the full history of the detected photons. There is a limitation as two events occurring with a delay smaller than \SI{100}{\nano \second} cannot be registered, thus creating an artificial flat region that has been removed from the signal for more clarity. A blue line indicates the $0.5$ threshold under which the emitter can be regarded as a single photon one. We can also note that the blinking of the emitter creates a bunching effect, that is clearly visible as the peaks at non-zero delay are higher than $1$: this kind of effect was previously observed with other types of emitters~\cite{manceau2018cdse, ulrich2005correlated}.  

\paragraph{Photostability} The main limit in the use of perovskytes nanocubes is due to their limited photostability: in particular, our emitters allowed measures for about \SI{15}{\min} before to be completely destroyed. The reasons of this is under investigation, but seems related to the simultaneous presence of moisture and light. A possible approach to partially solve this problem can be to protect them with a polymer. In particular it has been shown that polystyrene works efficiently to increase the spectral stability~\cite{raino2019underestimated}. Our preliminary results show that it can also improve the photostability by four times.
\section{Nanofibers}
\label{sec:nanofibers}

A nanofiber (\nanofiber) is a cylindrical glass optical waveguide with a diameter smaller than the wavelength of the guided light. It is fabricated by heating and pulling a standard single mode fiber in order to reduce its diameter up to the desired size. At the end of the process, in the \nanofiber{} part, the core and the cladding of the fiber are merged together and the light is coupled at the interface between the \nanofiber{} and the air surrounding it. An adiabatic transition between the unpulled fiber part and the \nanofiber{} region and vice-versa allows a good light transmission of the whole system and an easy way to inject the light to and collect the light from the \nanofiber{}.
The choice of the nanofiber profile is crucial to obtain a good coupling and a good transmission and the techniques has been largely studied \cite{stiebeiner_design_2010,nagai_ultra-low-loss_2014,birks_shape_1992}.

In a \nanofiber{}, the fiber modes are outside the fiber and are very intense near the surface \cite{kien_field_2004}: this property allows the light to interact with atoms or nanoparticles placed in the proximity of the surface. At the same time, it allows the light emitted by the nano-object to be coupled to the fiber~\cite{nayak2008single}, with a precise relation between the polarization of the emitted light and the polarization coupled inside the fiber~\cite{joos2018polarization}. 

For a step-index fiber to be single mode, the normalized operating frequency $V$ (also known as $V$-value) has to be smaller than the cutoff frequency $V_c=2.405$~\cite{kao1984fiber}. The following relation has hence to be valid: 
\begin{equation}
\label{eq:monomode}
V \equiv k a \sqrt{n_{1}^{2}-n_{2}^{2}}<V_{c} \cong 2.405
\end{equation}
where $a$ is the radius of the fiber, $n_1$ and $n_2$ are the two refractive indices of the core and the cladding and $k$ is the wavenumber of the coupled light. In our case $n_2=1$, as in a \nanofiber{} the role of the cladding is played by the air surrounding it. 
In addition, in order to couple a nanoobject, we want the light intensity $\left|E\right|^2$ to be the strongest in the vicinity of the surface. This is obtained when the diameter of the \nanofiber{} is smaller then the light wavelength, typically by a factor 2.

\begin{figure}
    \centering
    \includegraphics[width=0.5\linewidth]{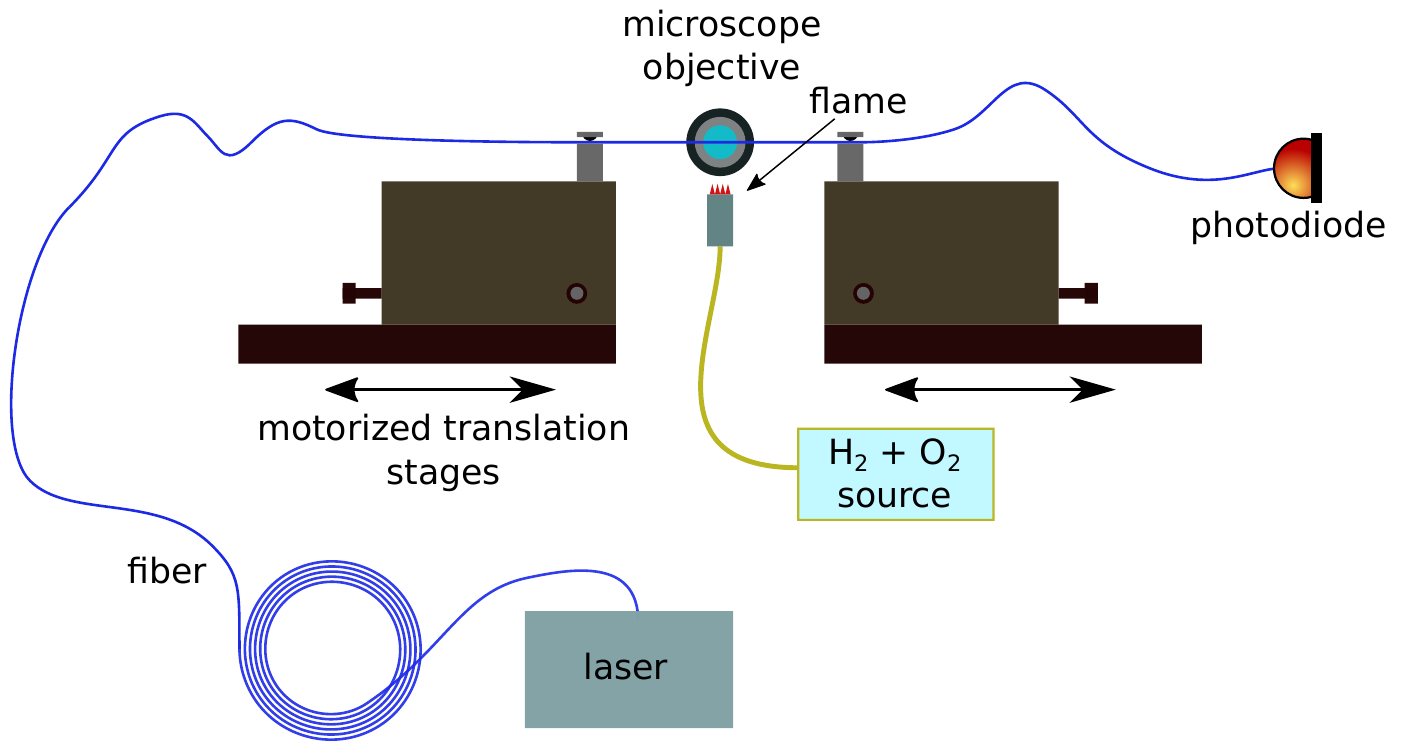}
    \caption{Scheme of the pulling setup. First the fiber is cleaned and clamped over the flame as shown, then the flame moves up and the two motorized platforms pull  the fiber and move it over the flame, while the procedure is imaged using a microscope objective on top. The transmission is monitored over the whole procedure by a photodiode at the output of the fiber. All the protocol is controlled by a computer.} 
    \label{fig:pulling_setup}
\end{figure}

The first step before pulling the fiber is to thoroughly clean it and remove the plastic jacket that covers the fiber. The fiber is then installed in the fabrication platform represented in figure~\ref{fig:pulling_setup}.
The fiber is suspended between two motorized translation stages that can move horizontally and are controlled by a computer. A Bunsen burner is carefully positioned under the fiber and another motor controls its vertical position, and provide a controlled flame of \ch{H2} and \ch{O2}. A photodiode, placed at the output of the fiber allows to monitor the transmission during the whole pulling process. In addition, the image of the fiber is collected by a microscope objective and recorded by a camera behind it: this is useful for a correct positioning of the flame and to precisely adjust its vertical position during the pulling. 

When the pulling starts, the flame is placed at a given distance from the fiber and the motors start moving and pulling the fiber in order to obtain the desired shape. 
The movements needed to perform this operation depend not only on the chosen profile, but also on the physical characteristics of the flame, such as its size and temperature. The distance between the flame and the fiber, as well as the flux of hydrogen and oxygen need to be carefully chosen in order to obtain the effective flame diameter (in our case \SI{0.5}{\milli\meter}) used to calculate the fiber profile. 
The effect of the flame size in the pulling procedure is described in detail in~\cite{hoffman2014optical}.
With this system, we were able to reach fiber radii as small as \SI{150}{\nm} and to observe transmissions over $95\%$. By accurately optimizing the shape of the tapered part even higher transmissions, approaching unity, can be achieved~\cite{nagai_ultra-low-loss_2014}.

\paragraph{Coupling of single emitters}
Nanocristals can be coupled to the nanofiber platform by exploiting the strong evanescent field present at the nanofiber surface. Indeed, a nanocrystal placed on the nanofiber surface will emit some of the single photons directly inside the nanofiber: this is an easy and compact way to achieve fiber-integrated single photons sources.

In order to place a nanoemitter on the nanofiber, we prepare a drop of solution of \SI{20}{\micro\liter} containing few nanoemitters at the end of a micro-pipette. The next step is then to touch the \nanofiber{} with the drop repetitively until a particle gets stuck to it. For more precision, the whole process is controlled with a microscope and the micro-pipette is moved via micrometric translation stages.

\paragraph{Experimental apparatus}

\begin{figure}
    \centering
    \includegraphics[width=0.8\linewidth]{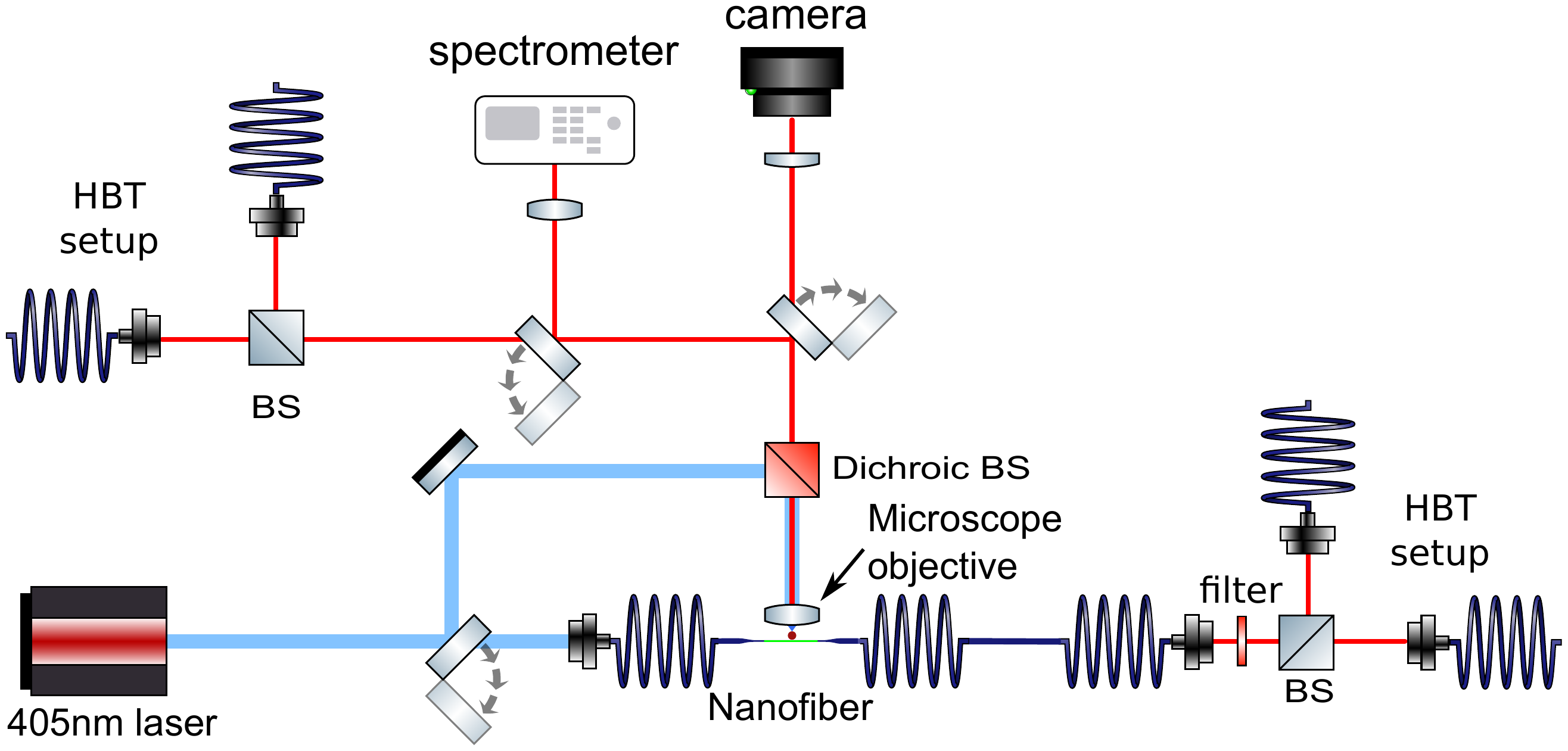}
    \caption{Scheme of the optical setup used to measure the light emitted by nanoemitters deposited on the \nanofiber{}. The green line show the path of the excitation laser. Changing the position of the beamsplitter we can send it through our confocal microscope or directly at the end of the \nanofiber{}. The red line shows the detection path, that allows to measure alternatively the image of the emitter, its spectrum or the $g^2(\tau)$ function.}
    \label{fig:fiber_setup}
\end{figure}

The setup is represented in figure~\ref{fig:fiber_setup}. By sending the light in the fiber we can excite the emitter and detect its emission through the microscope: this is very convenient to detect the emitted light without having to care about the precise position of the emitter and to verify its single photon emission. If the coupling is good enough, we can also excite the nanoemitter through the microscope and collect the light from the fiber.

First results were obtained on our set-up with semiconductor nanocrystals that are well known to be good single photon emitters~\cite{michler2000quantum} and so well adapted for first tests. 
The emitters we used are dot-in-rod \ch{CdS/CdSe} nanocrystals: this kind of nanocrystals have a core/shell structure which ensures a reduced blinking of the emission;
moreover dots in rod have the specificity to have elongated shell that makes the emission polarized~\cite{pisanello2010room}.

\begin{figure}
    \centering
    \includegraphics[width=0.5\linewidth]{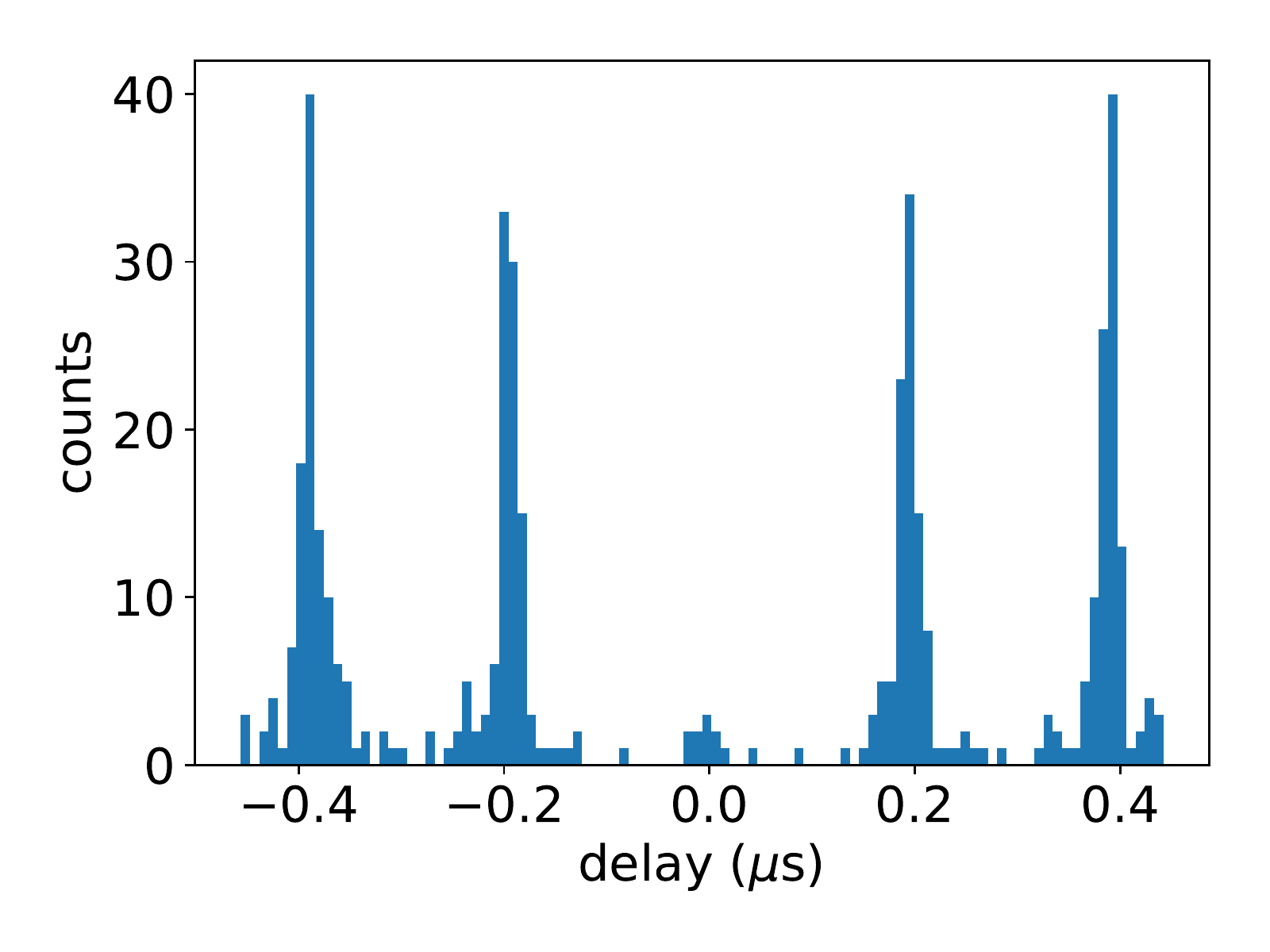}
    \caption{$g^{(2)}(\tau)$ function measured by exciting a dot-in-rod nanocrystal deposited on the \nanofiber{} with a \SI{405}{\nm} pulsed laser. The repetition rate was \SI{5}{\mega\hertz}. The photons were collected via the \nanofiber{}. }
    \label{fig:g2_from_fiber}
\end{figure}

For that, we pulled a fiber with a diameter of \SI{300}{\nm} in order to guide the \SI{600}{\nm} light emitted by the nanocrystals. As a result, we were able to excite a single nanocrystal from the free space. The emitted light collected by the \nanofiber{} was enough to measure the photon autocorrelation function ($g^{(2)}(\tau)$) using a Hanbury Brown and Twiss set-up (figure~\ref{fig:g2_from_fiber}). For this experiment a pulsed excitation laser with a wavelength of \SI{405} {\nm} was used. A clear signature of a single photon emitter was observed with a $g^{(2)}(\tau)$ smaller than $0.2$. This first result validates our experimental protocol for the use of other emitters and confirm that we can use different emitters in future experiments such as perovskite nanocrystals or silicon-vacancy centers in nanodiamonds.

\section{Conclusion \& perspectives}
In this article, we presented a nanophotonics platform for efficient coupling of light with nanoemitters. We also presented recent novel results on the optical properties of perovskite nanocrystals, including quantum optical properties. We finally explored a way towards a fully integrated platform where light from nanoemitters is emitted directly into an optical fiber.

To fully exploit the potential scalability of the nanofiber-based platform many development remain to be done. In the future, achieving a deterministic positioning of the nanoemitters on the nanofiber could open the way to more complex structures, such as cavities made by laser-writing Bragg mirrors on the fiber or nanofiber-resonators to improve the spectral properties of the emitted light.
 
\ack
The authors would like to acknowledge the funding from ANR projects Iper-nano2. AB et QG are members of the "Institut Universitaire de France" (IUF).

\clearpage

\bibliography{references}

\end{document}